# Diffusive hydrogenation reactions of CO embedded in amorphous solid water at elevated temperatures ~70 K

Masashi Tsuge*, Hiroshi Hidaka, Akira Kouchi, Naoki Watanabe

Institute of Low Temperature Science, Hokkaido University, Sapporo 060-0819, Japan

*Author to whom correspondence should be addressed

Tel: +81-11-7065474

E-mail: tsuge@lowtem.hokudai.ac.jp

Postal address: Kita-19, Nishi-8, Kita-ku, Sapporo 060-0819, Japan

Short title: Diffusive hydrogenation of CO embedded in ASW

# ABSTRACT

The surface processes on interstellar dust grains have an important role in the chemical evolution in molecular clouds. Hydrogenation reactions on ice surfaces have been extensively investigated and are known to proceed at low temperatures mostly below 20 K. In contrast, information about the chemical processes of molecules within an ice mantle is lacking. In this work, we investigated diffusive hydrogenation reactions of carbon monoxide (CO) embedded in amorphous solid water (ASW) as a model case and discovered that the hydrogenation of CO efficiently proceeds to yield $H_2CO$ and $CH_3OH$ even above 20 K when CO is buried beneath ASW. The experimental results suggest that hydrogen atoms diffuse through the cracks of ASW and have a sufficient residence time to react with embedded CO. The hydrogenation reactions occurred even at temperatures up to ~70 K. Cracks collapse at elevated temperatures but the occurrence of hydrogenation reactions means that the cracks would not completely disappear and remain large enough for penetration by hydrogen atoms. Considering the hydrogen-atom fluence in the laboratory and molecular clouds, we suggest that the penetration of hydrogen and its reactions within the ice mantle occur in astrophysical environments.

*Unified Astronomy Thesaurus concepts:* Astrochemistry (75); Molecular clouds (1072); Dense interstellar clouds (371); Interstellar molecules (849); Interstellar dust (836); Laboratory astrophysics (2004);

# 1. INTRODUCTION

Compared with surface processes that involve H atoms on astronomical dust grains, chemical evolution within ice mantles is unclear. To elucidate the beneath-mantle processes that involve H atoms impinging on astronomical ice, information about the penetration and diffusion of H atoms through water ice is important because water is the most abundant component of an ice mantle (Boogert et al. 2015) in dense molecular clouds. Recently, Minissale et al. (2019) experimentally investigated the penetration and diffusion of oxygen and deuterium atoms into amorphous solid water (ASW) by monitoring the reaction of atoms with nitrogen monoxide (NO) molecules embedded in ASW. For the thickest ice (10-monolayer ASW on 1-monolayer NO), the consumption of tracer NO after D-atom irradiation at 10 K only increased to the 0.1 monolayer. The authors concluded that surface hydrogenation is dominant over processes that involve penetration and diffusion in ice mantles. Awad et al. (2005) estimated the diffusion constant of hydrogen atoms into $H_2O$-CO mixed ice to be in the range $10^{-21}$–$10^{-20}$ cm$^2$ s$^{-1}$ at temperature range 10–15 K and concluded that H-atom diffusion into ice is significantly slower than surface diffusion. Petrik et al. (2014) explored low-energy (100 eV) electron-stimulated reactions in layered $H_2O$/CO ice. The H atoms produced within ASW diffused through the ASW and reacted with deeply buried CO molecules to yield products at temperature above ~60 K. Electron-induced processes in a dense molecular cloud are considered inefficient; e.g., at 10 K, electrons collide with dust grains once every 25000 years (Mason & Field 2019), which is much less frequent than H atoms that strike once a day (Watanabe & Kouchi 2008). Therefore, processes that involve nonenergetic (thermal) H atoms from the gas phase have particular importance for dense molecular clouds.

In theoretical chemical models, which are referred to as three-phase models, the grain surface and mantle are treated as separate phases (e.g., Hasegawa & Herbst 1993). Following Hasegawa & Herbst (1993), models that include active bulk chemistry have also been developed (Kalvāns & Shmeld 2010; Garrod 2013; Ruaud et al. 2016). Inclusions of UV-induced chemistry, which accompanies the diffusive reaction of radicals, within molecular mantles could reproduce the observed ice species in some environments (Ruaud et al. 2016). However, the diffusion rates of lighter elements (e.g., H, $H_2$, C, N, and O) within a mantle were assumed arbitrarily without experimental justifications.

The successive hydrogenation of CO has been most investigated as a molecular formation process on ice because its products, formaldehyde ($H_2CO$) and methanol ($CH_3OH$), were abundantly detected in astronomical ice (Gibb et al. 2004) and some comets (Crovisier & Bockelée-Morvan 1999). In dense molecular clouds, the successive hydrogenation of CO on grains has been adopted as the most plausible pathway to form these organic molecules over photolysis (Allamandola et al. 1988; Schutte et al. 1996), and/or cosmic-ray or proton bombardment of $H_2O$-CO ice. Watanabe & Kouchi (2002) showed that $H_2CO$ and $CH_3OH$ are efficiently produced on the surface of $H_2O$-CO ice at 10 K, where the fluence of H atoms was relevant to dense molecular clouds. Subsequently, CO hydrogenation reactions on various ice compositions (pure CO, CO-capped $H_2O$, and $H_2O$-CO mixed ice) at various temperatures were investigated (Watanabe et al. 2003; Watanabe et al. 2004; Fuchs et al. 2009). One important finding of these studies is that the efficiency of CO hydrogenation reactions on ice significantly drops near 20 K probably due to the reduction in the coefficients for H-atom sticking to ice. As a result, the successive CO hydrogenation has generally been assumed to be valid on the ice surface only at temperatures approximately below 20 K.

In this work, we employed CO molecules as a tracer to investigate whether H atoms can reach molecules that are buried beneath ASW and further cause their hydrogenation. We performed experiments at various temperatures from 10 to 70 K with an ASW thickness that ranged from 1 to 80 monolayers. The hydrogenation of CO efficiently occurred even at relatively high temperatures ($\geq$ 20 K) when CO molecules were embedded underneath thicker ASW. Considering the dependency on temperature and ASW thickness, we discuss the mechanisms that lead to efficient hydrogenation reactions. In addition, experiments with D atoms indicated that the H(D)-atom diffusion within ASW is dominated by thermal hopping.

## 2. EXPERIMENT

Experiments were performed using the apparatus named LAboratory Setup for Surface reactions in Interstellar Environments (LASSIE), which has been described elsewhere (Watanabe et al. 2003; Hidaka et al. 2004). The apparatus consists of an ultrahigh vacuum main chamber that is equipped with a liquid nitrogen shroud and a differentially pumped atomic hydrogen source chamber. A polished aluminum substrate is located in the center of the main chamber and can be cooled to 9 K by a closed cycle helium cryostat (RDK-408R, SHI). Ice samples were made by background deposition with the following procedures. First, approximately 5 monolayers (MLs) of CO were deposited onto the substrate at 9 K with an approximate deposition rate of 1 ML min$^{-1}$. In this condition, CO solid forms an amorphous structure, while the CO solid transforms into a crystalline structure at temperatures above approximately 20 K. Therefore, to produce the same CO structure in the experiments, the temperature of the substrate was once raised to 20 or 25 K before water deposition. ASW was produced over the CO solid kept at 20 or 25 K with an approximate deposition rate of 1 ML min$^{-1}$. The ice-thickness was estimated from integrated absorbance with reported integrated absorption coefficients; $1.1 \times 10^{-17}$ and $2.0 \times 10^{-16}$ cm molecule$^{-1}$ for CO (Jiang et al. 1975) and H$_2$O (Hagen et al. 1981), respectively. For CO solid and ASW, the number density which comprises 1 ML is assumed to be $1 \times 10^{15}$ molecules cm$^{-2}$. Islands are formed when crystalline CO (α-CO) is thermally produced from amorphous CO (Kouchi et al. in prep.). Although the layered structure, i.e., ASW layer on a flat CO surface, might not be valid especially at elevated temperatures, we refer to the samples as layered ice for simplicity. The temperature of the layered ice was then set within the range 10–70 K for H-atom irradiation. Hydrogen atoms were generated by dissociating H$_2$ gas in a microwave-induced plasma in a Pyrex tube and transferred via a sequence of PTFE and aluminum tubes onto the ice sample. The aluminum tube was connected to another cryostat and cooled to approximately 100 K. The kinetic energy of impinging H atoms is thought to be thermalized to the temperature of Al pipe (Nagaoka et al. 2007). The dissociation fraction of the impinging beam was approximately 20%. The flux of the atomic beam at the sample surface was estimated to be $\sim 1 \times 10^{14}$ atoms cm$^{-2}$ s$^{-1}$ according to Hidaka et al. (2007), where a relation between the flux and the effective rate constant for the CO decay for a CO-capped ASW ice at 15 K was presented. The H-atom flux was stable within the duration of experiment and experiment-to-experiment. The ice samples were monitored in situ by a Fourier transform infrared spectrometer with a resolution of 4 cm$^{-1}$.

## 3. RESULTS AND DISCUSSION

Figure 1A shows the infrared (IR) difference spectra measured for pure CO and layered ASW/CO ice at 20 K irradiated with H atoms for 180 min (corresponds to H-atom fluence ~1 × $10^{18}$ atoms cm$^{-2}$); positive bands and negative bands indicate generation and destruction, respectively. For the layered ice, the IR bands of $H_2CO$ were observed at 1724, 1502, and 1247 cm$^{-1}$, and that of $CH_3OH$ was observed at 1031 cm$^{-1}$, while a distinct decrease of CO was observed near 2153 cm$^{-1}$. In contrast, only a small amount of $H_2CO$ was identified for pure CO ice at 20 K, which is consistent with that reported in the literature (Watanabe et al. 2004; Hidaka et al. 2007; Fuchs et al. 2009).

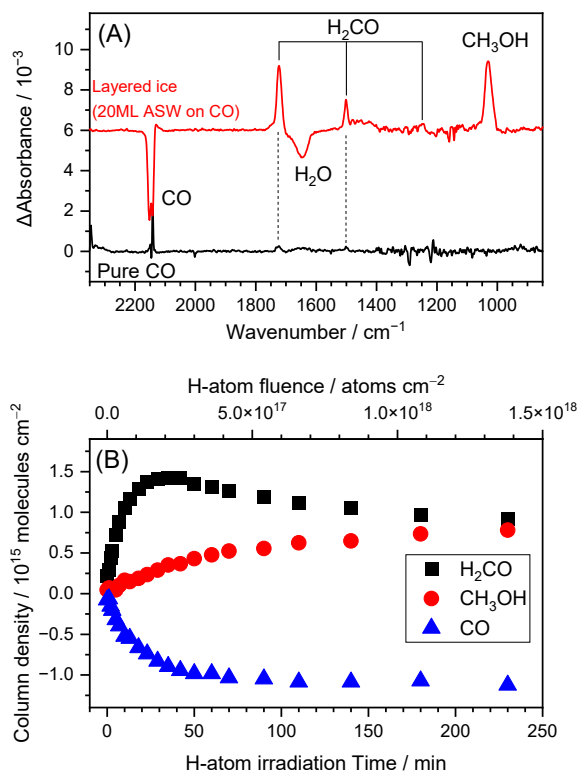

Figure 1. (A) Infrared difference spectra that shows the results of 180-min H-atom irradiation on pure CO solid (lower trace) and a layered ice with 20 MLs ASW on CO (upper trace). The layered ice was made by depositing water at 20 K over a CO solid. H-atom irradiation was conducted at 20 K for both ice samples. The infrared bands due to CO, $H_2CO$, $CH_3OH$, and $H_2O$ are indicated. (B) Variations in column densities (in molecules cm$^{-2}$) for layered ice (20 MLs ASW on CO) with H-atom irradiation time. The H-atom fluence, which is calculated assuming an H-atom flux of 1 × $10^{14}$ atoms cm$^{-2}$ s$^{-1}$, is given at the top axis. The statistical error is within the data symbols.

The time variations of the column densities of CO, $H_2CO$, and $CH_3OH$ are shown in Figure 1B for the layered ice irradiated at 20 K. The column densities were calculated using integrated absorbance and integrated absorption coefficients reported for pure solids; $1.1 \times 10^{-17}$, $9.6 \times 10^{-18}$, and $1.8 \times 10^{-17}$ cm molecule$^{-1}$ for CO (2153 cm$^{-1}$; Jiang et al. 1975), $H_2CO$ (1724 cm$^{-1}$; Schutte et al. 1993), and $CH_3OH$ (1031 cm$^{-1}$; Kerkhof et al. 1999), respectively. The sum of the column densities for $H_2CO$ and $CH_3OH$ exceeds the decrease in the column density for CO probably due to uncertainties in the estimation of the column densities. The observed temporal variation is similar to that reported for $H_2O$-CO mixed ice below 15 K (Watanabe et al. 2003; Watanabe et al. 2004), which indicates that the successive hydrogenation of CO occurs in the layered ice even at 20 K. The decrease in CO column density was saturated at approximately $1 \times 10^{15}$ molecules cm$^{-2}$ after 60–100 min irradiation, while the conversion of $H_2CO$ to $CH_3OH$ was still proceeding.

A series of experiments were performed by varying the thickness of the ASW layer from 1 to 80 MLs. Figure 2A shows the temporal variations in the column densities of CO, $H_2CO$, and $CH_3OH$ measured for layered ice with (from bottom to top) 1, 6, 10, and 40 MLs ASW on solid CO. The decreased CO column densities after 5, 20, and 200 min H-atom irradiation, which corresponds to a fluence of $2 \times 10^{16}$, $1 \times 10^{17}$, and $1 \times 10^{18}$ atoms cm$^{-2}$, respectively, are plotted in Figure 2B. For the ASW thickness of 1–6 MLs, the CO consumption increased very slowly and continued even after 200 min H-atom irradiation. When the thickness exceeds 10 MLs, a significant increase in the CO consumption was observed. We will discuss this sudden change in reactivity later. For the thicknesses of 10, 20, 40, and 80 MLs, the CO consumption approached the asymptotic value of approximately $1.1 \times 10^{15}$ molecules cm$^{-2}$ after 60–100 min irradiation, as shown in Figures 1B and 2A. The rate and asymptotic values for CO consumption are similar for thicker layered ice; therefore, the diffusion of H atoms within ASW layers is rapid within the experimental time scale.

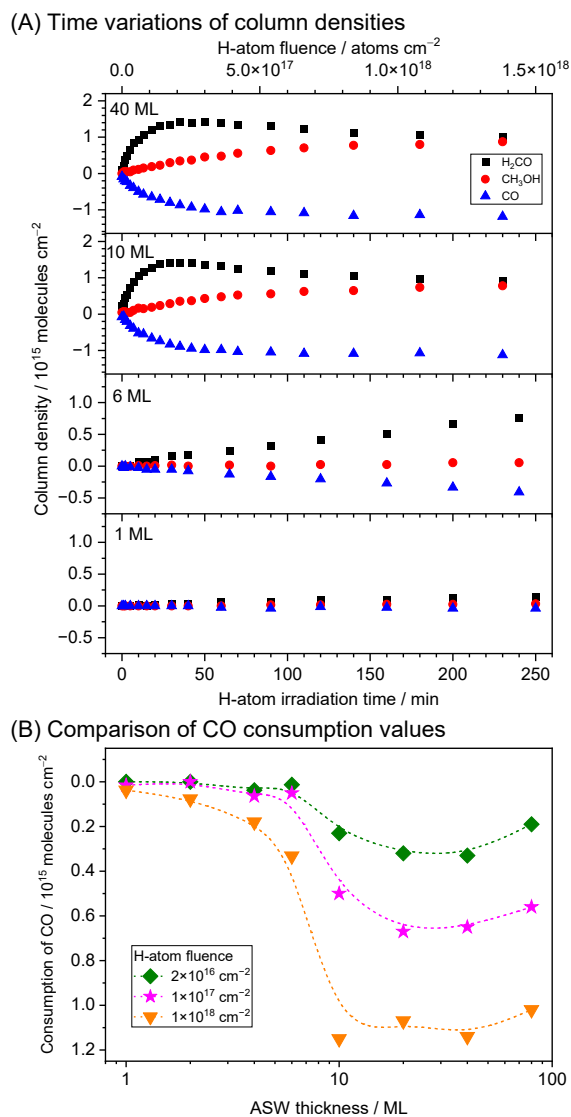

Figure 2. Dependence on ASW thickness of the CO hydrogenation reaction. (A) Variations in column densities (in molecules cm$^{-2}$) for layered ice (1–40 MLs ASW on CO) with H-atom irradiation time. H-atom irradiation was conducted at 20 K. The H-atom fluence calculated assuming the H-atom flux of $1 \times 10^{14}$ atoms cm$^{-2}$ s$^{-1}$ is given at the top axis. (B) The ASW thickness dependence of the CO consumption values investigated at 20 K for layered ice 1–80 MLs ASW on 5 MLs CO. The CO consumption values after 5, 20, and 200 (or 180) min irradiation, which corresponds to a fluence of $2 \times 10^{16}$, $1 \times 10^{17}$, and $1 \times 10^{18}$ atoms cm$^{-2}$, respectively, are plotted by different symbols. Dashed lines (B-spline) are shown as eye guides.

Another series of experiments were performed to investigate the temperature dependence. These experiments were performed with a layered ice of 20 MLs ASW on 5 MLs CO and a temperature range 10–70 K. In Figure 3A, the temporal variations of the column densities of CO, $H_2CO$, and $CH_3OH$ measured at temperatures (from bottom to top) 10, 20, 30, and 50 K are shown. Although $CH_3OH$ was not observed at 50 K, there is a weak scattering in the estimated column density of $CH_3OH$. The scattering that amounts to $< 1 \times 10^{13}$ molecules cm$^{-2}$ is representative of the degree of error from noise. Therefore, we would say that the minimum detectable change of column density is $(1–2) \times 10^{13}$ molecules cm$^{-2}$ and the estimated abundance of $H_2CO$ at 50 K, up to $9 \times 10^{13}$ molecules cm$^{-2}$, is more significant than the noise level. The CO consumption values after H-atom irradiation for a fluence of $2 \times 10^{16}$, $1 \times 10^{17}$, and $1 \times 10^{18}$ atoms cm$^{-2}$, are plotted in Figure 3B. In the temperature range 10–20 K, the CO consumption increased as the temperature increased. The observed trend would be attributed to the thermal activation of H-atom diffusion which enables H atoms to reach CO molecules underneath ASW. The efficiency of CO-hydrogenation reactions in CO-capped ASW samples or $H_2O$/CO mixed ice is known to be significantly reduced at temperature above ~20 K (e.g., Watanabe & Kouchi 2008). Therefore, the scenario that the observed hydrogenation occurs in the vicinity of ASW surface followed by CO diffusion from the bottom of ASW can be ruled out. In addition, it is quite reasonable to consider that the H-atom diffusion is much faster than that of CO molecule. Nevertheless, once H atoms reach the deep area of ASW, CO diffusion over relatively short distance may enhance the chance for encountering H atoms. The CO consumption value gradually decreased as the temperature increased from 20 K. Because H-atom diffusion should become more efficient at elevated temperatures, the decrease would correspond to the drop in the number density of the H atoms in ASW. We note that, above 40 K, the layered structure might not be retained because CO molecules would partially sublime. We will discuss the structural changes and reaction mechanisms in a subsequent section.

The bulk diffusion of H atoms in ice $I_h$ is unmeasurably slow at temperatures below 40 K (Markland 2008). In our experiments, since the hydrogenation reactions were activated even in the lower temperature range of 10–12 K, the diffusion of H atoms within ASW occurs through cracks.

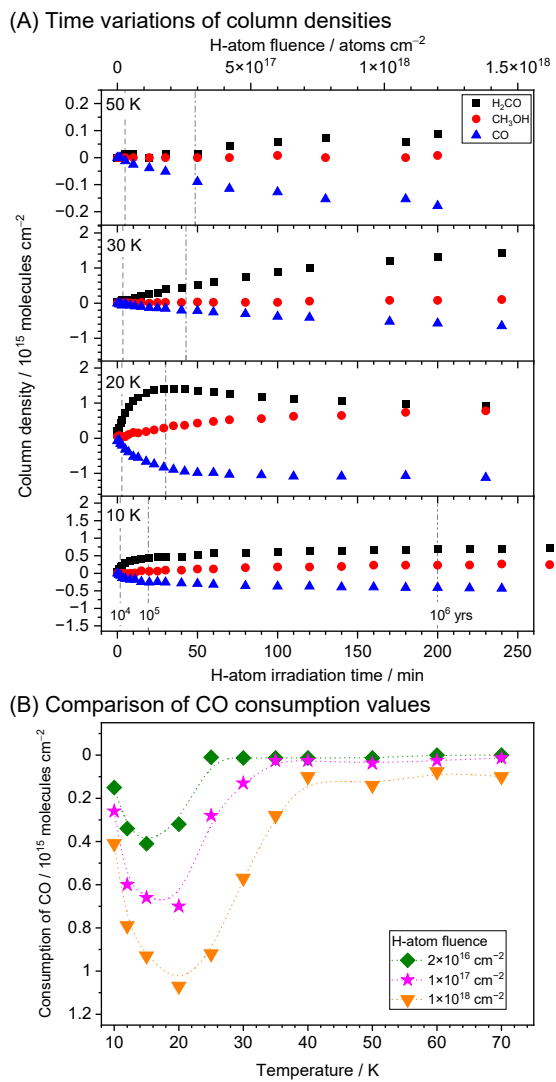

**Figure 3.** Temperature dependence of the CO hydrogenation reaction. (A) Variation in the column density (in molecules cm$^{-2}$) for layered ice (20 MLs ASW on CO) with H-atom irradiation time at temperatures (from bottom to top) of 10, 20, 30, and 50 K. The H-atom fluence, which is calculated assuming the H-atom flux of $1 \times 10^{14}$ atoms cm$^{-2}$ s$^{-1}$, is given at the top axis. The fluence of H atom in molecular clouds in $10^4$, $10^5$, and $10^6$ years are indicated with gray lines (dash, dash-dot, and dash-dot-dot, respectively). These values were calculated assuming that the number density of H atom is 1 cm$^{-3}$ and the temperatures of gas and dust are the same. (B) Temperature dependence of the CO consumption values investigated for layered ice with 20 MLs ASW on 5 MLs CO in the temperature range 10–70 K. The CO consumption values after 5, 20, and 200 (or 180) min irradiation, which corresponds to a fluence of $2 \times 10^{16}$, $1 \times 10^{17}$, and $1 \times 10^{18}$ atoms cm$^{-2}$, respectively, are plotted by different symbols. Dashed lines (B-spline) are shown as eye guides.

Figure 4 shows the IR reflection absorption spectra of the CO band at 2143 cm$^{-1}$ for the CO solid deposited at 9 K. No significant change in the spectra was observed after warming the pure CO solid sample to 20 or 25 K. After deposition of 20 MLs of ASW on the CO solid, the CO band was split into two components: a stronger peak at 2138 cm$^{-1}$ and a weaker peak at 2153 cm$^{-1}$. For $H_2O$-CO mixed ice, the former is attributed to CO molecules in a nonpolar environment, and the latter is attributed to those in a polar environment (Bouwman et al. 2007). Another interpretation is found in literature (e.g., Fraser et al. 2004); the peak at 2153 cm$^{-1}$ is due to CO on dangling OH sites of ASW while the peak at 2138 cm$^{-1}$ is partly due to CO adsorbed sites without dangling OH and partly due to pure CO ice. As shown at the bottom of Figure 4A, irradiation of layered ice with H-atoms at 20 K induced the predominant decrease of the 2153 cm$^{-1}$ component, indicating that reacting CO is in a polar environment, i.e., at the ASW/CO interface. When the layered ice sample was warmed to 50 K, partial sublimation to 3 MLs of CO was observed (see the second trace from the bottom in Figure 4B). The spectrum at 50 K shows one intense peak at 2138 cm$^{-1}$, while the peak at 2153 cm$^{-1}$ mostly diminishes. As will be subsequently discussed, we deduce that this variation reflects the disappearance of CO molecules from the surface of ASW cracks due to sublimation and/or formation of **CO aggregates**.

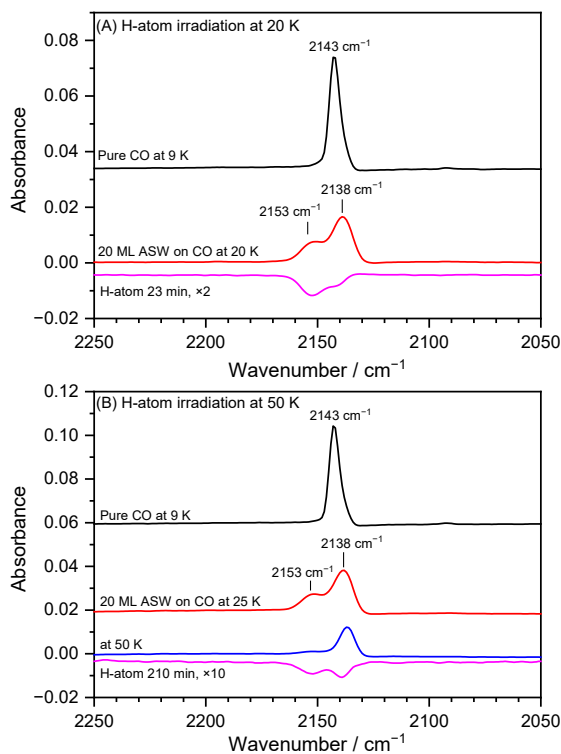

**Figure 4.** Infrared spectra in the CO region. (A) IR spectra measured in the experiment of H-atom irradiation at 20 K. Top, IR spectrum of pure solid CO deposited at 9 K; middle, IR spectrum measured after deposition of 20 MLs ASW over solid CO at 20 K; bottom, the difference spectrum that shows the result of H-atom bombardment at 20 K for 23 min. (B) IR spectra measured in the experiment of H-atom irradiation at 50 K. Top, the IR spectrum of pure solid CO deposited at 9 K; second top, IR spectrum measured after deposition of 20 MLs ASW over solid CO at 20 K; second bottom, IR spectrum of the layered sample measured at 50 K; bottom, the difference spectrum that shows the result of H-atom bombardment at 50 K for 210 min.

These experimental observations are discussed using the schematics of the ice samples shown in Figure 5. The deposition of CO at 9 K produces amorphous CO; the thickness is estimated to be 5 MLs (Figure 5A). Crystallization of solid CO occurs at 17–25 K to produce α-CO, which forms islands with a height of approximately 15 MLs, as observed by a transmission electron microscope (Kouchi et al. in prep.) (Figure 5B). As shown in Figure 5C, because of island formation, a thin ASW layer will not fully cover the CO layer. This situation would correspond to layered ice with an ASW thickness less than 6 MLs, where the CO consumption values were relatively small (Figure 2B). In this case, only CO molecules exposed to a vacuum have a chance to be hydrogenated, similar to H-atom irradiation on $H_2O$-CO mixed ice and CO-capped $H_2O$ ice. When the thickness of ASW increases (Figure 5D), the CO layer is eventually covered with ASW. Because the ASW layer was produced at low temperatures, 20 or 25 K, it is porous and has cracks. At these temperatures, CO can partly diffuse along the surface of an ASW crack (Mispelaer et al. 2013; Karssemeijer et al. 2014; Lauck et al. 2015; He et al. 2018), which causes adsorption of CO on the walls of cracks, as depicted in Figure 5D. This adsorbed CO molecule and the molecules at the ASW/CO-island interface are regarded as CO molecules in a polar environment. CO molecules within islands can be regarded as those in a nonpolar environment. The existence of these two types of environments is supported by IR spectra (Figure 4). Hydrogen atoms that penetrate an ASW layer through cracks will have a chance to react with CO molecules. H atoms will be trapped in cracks especially when bottlenecks exist along cracks, leading to a long residence time.

When the temperature of a sample is raised above the sublimation temperature of CO, a significant decrease in the CO absorption intensity was observed (see Figure 4). At these temperatures, efficient diffusion of CO occurs through ASW cracks. However, at temperatures below 70 K, a fraction of CO remained, and the decreased amount was as large as 3 MLs out of 5 MLs. We deduce that the remaining CO molecules produce aggregates of CO that are trapped in ASW. We note that the precise molecular configuration of CO trapped in ASW has been considered uncertain (e.g., Collings et al. 2003 & Fraser et al. 2004). Further study to determine the CO structure would be necessary in the future. The trapping of CO at elevated temperature occurs due to the collapse of cracks as shown in Figure 3 of Collings et al. (2003); this situation is depicted in Figure 5E. From the experimental observations that some CO consumptions were observed even at 70 K, it is anticipated that H atoms can still diffuse through narrow cracks. The significant decrease in CO consumption above 40 K (Figure 3B) might be explained as follows: above 40 K, the sticking (physisorption) of H atoms at the ASW/vacuum interface hardly occurs, and therefore, only H atoms that directly impinge the cracks can penetrate ASW and eventually react with CO. Furthermore, even when H atoms reach CO molecules, the interaction time may not be enough to cause a hydrogenation reaction. Although the surface area and density of ASW largely differ between 40 K and 70 K (Stevenson et al. 1999; Brown et al. 1996), the CO consumption values are similar at temperatures above 40 K; this result suggests that the effect of the width of cracks on H-atom diffusion is limited.

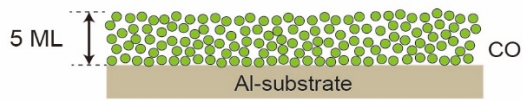
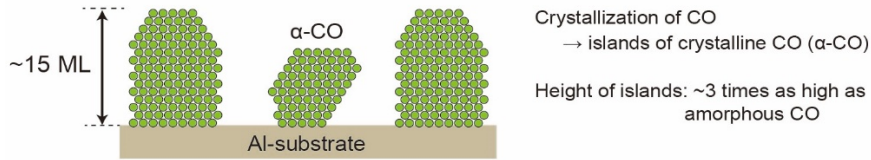
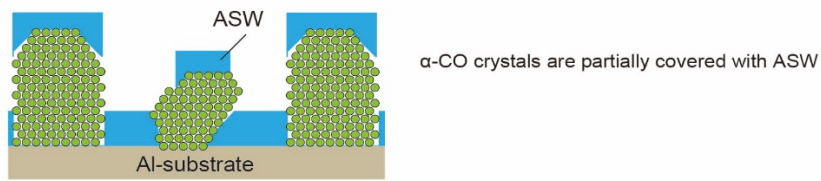
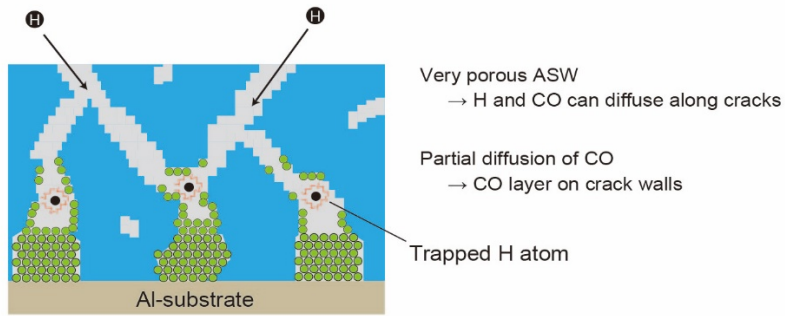
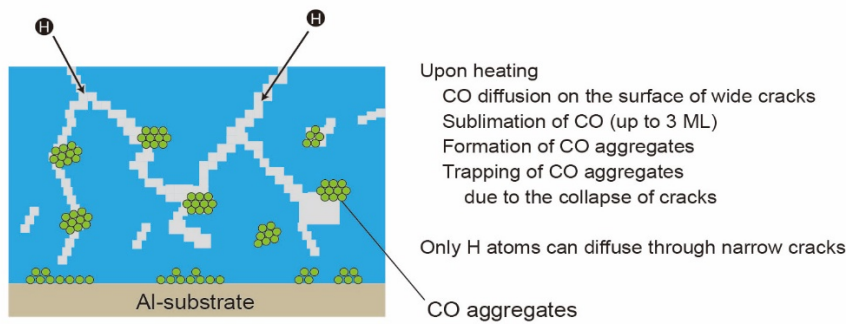

**Figure 5.** Schematic of the ASW/CO layered ice investigated in this study (not to scale).

Previous experimental studies indicated that ASW produced by surface reactions at low temperatures is predominantly nonporous (e.g., Oba et al. 2009). Unfortunately, preparation of layered ice, which comprises nonporous ice on solid CO, is impossible because of the volatility of CO. To gain further insight into H-atom penetration and diffusion in astronomically relevant ice, experiments with a less volatile molecule that easily reacts with H atoms are underway. We discovered that H atoms efficiently penetrate an ASW-covered ice annealed at 70 K, for which IR absorptions due to OH-dangling bonds were not observed (Oba et al. 2009).

To investigate the mechanism of diffusion, thermal hopping or tunneling-mediated, experiments were performed for a layered ice at 20 K using D atoms instead of H atoms. When a layered ice with 20 MLs ASW on CO solid was irradiated with D atoms, the formation of $D_2CO$ and $CD_3OD$ and a decrease in CO were observed at slower rates than hydrogenation. However, these slower rates can be attributed to the isotope effect in H + CO and H + $H_2CO$ reactions (Hidaka et al. 2007; Watanabe & Kouchi 2008), which proceed via quantum mechanical tunneling due to large barriers (e.g., Woon 1996; Woon 2002). In addition, we performed experiments using another type of layered ice: 20 MLs ASW on $O_2$. The successive hydrogenation reactions of $O_2$ that generate $H_2O_2$ are barrierless processes, and thus, no isotope effect is expected (Miyauchi et al. 2008). For H-atom-irradiated layered ice and D-atom-irradiated layered ice (20 MLs ASW on $O_2$), the formation rates for $H_2O_2$ and $D_2O_2$, respectively, were similar. This result indicates no significant isotope effect in the diffusion of H(D) atoms; therefore, we deduce that the H-atom diffusion through the cracks of ASW is dominated by a thermal hopping mechanism, similar to H-atom diffusion on the ASW surface (Hama et al. 2012). In diffusive hydrogenation reactions, hydrogen atoms need to diffuse a long distance. It has been reported that such long-distance diffusion of hydrogen atoms on the ASW surface is dominated by thermal hopping (Kuwahata et al. 2015). As mentioned earlier, H (D) atoms need to diffuse (penetrate) deep into the cracks of ASW to attain a sufficient residence time for hydrogenation reactions. The absence of kinetic isotope effect in the hydrogenation of $O_2$ underneath ASW requires condition that the number density of trapped atoms at reaction sites is similar in H-atom and D-atom irradiation experiments. Therefore, even if the reaction between trapped H (D) atoms and $O_2$ molecules occur upon a short distance diffusion of $O_2$, the absence of kinetic isotope effect can only be attributed to the thermal diffusion of H (D) atoms.

In this work, we discovered that the CO hydrogenation reaction to produce $H_2CO$ and $CH_3OH$ occurs after H-atom irradiation of ASW/CO layered ice. The observation of products at elevated temperatures, i.e., above 20 K, is important because the hydrogenation of CO on the surface of ice was known to be inefficient at these temperatures (Watanabe & Kouchi 2008). In our experiments, H-atom flux and duration of irradiation can cause fluence that is relevant to dense molecular cloud environments. Assuming that the number density of H atoms is 1 cm$^{-3}$, the fluence of hydrogen in a 10 K molecular cloud becomes $1.3 \times 10^{16}$, $1.3 \times 10^{17}$, and $1.3 \times 10^{18}$ atoms cm$^{-2}$ over $10^4$, $10^5$, and $10^6$ years, respectively; the fluence approximately corresponds to 2, 20, and 200 min irradiation in the experimental conditions of this study. Therefore, we suggest that $H_2CO$ and $CH_3OH$, which are the abundant components of astronomical ice, can be formed not only on the surface but also within ASW in astrophysical conditions.

The dependence on the ASW thickness indicates that efficient hydrogenation reactions occur when the CO surface is fully covered with ASW (Figure 5D). From the identification of polar and nonpolar CO molecules in astronomical environments (Sandford et al. 1988; Tielens et al. 1991), we deduce that this situation can be realized in molecular clouds, i.e., ASW/CO interface in a layered ice (onion-like structure) and CO aggregates embedded in ASW. The formation of the CO aggregates in ASW has been inferred from many experimental studies (e.g., Collings et al. 2003). Therefore, we suggest that diffusive hydrogenation reactions within an ice mantle would be more effective than ever considered, although further verifications by theoretical chemical models are required to determine their impact on the chemical evolution. In models, the sticking of H atoms to a surface and their penetration are treated as a sequential process; impinging H atoms that are directly trapped within cracks of ice above 20 K are not considered, which underestimates the H-atom related processes within an ice mantle. Our findings indicate that H atoms penetrate ASW after collision with dust and are trapped for a sufficient period leading to hydrogenation reactions.

## Acknowledgments

This work was supported by the Japan Society for the Promotion of Science (JSPS KAKENHI Grant Nos. JP18K03717 and JP17H06087).